\documentclass[8pt]{article}
\renewcommand{\baselinestretch}{1.5}
\usepackage{sectsty}
\usepackage[pdftex]{graphicx}
\usepackage{bm}
\usepackage{color}
\usepackage{textcase, url}
\usepackage[normalem]{ulem}
\usepackage[colorlinks=true, linkcolor=black, urlcolor=black, citecolor=black]{hyperref}
\usepackage{amsmath,amssymb}
\usepackage{mathtools}
\usepackage{lineno}
\usepackage{tabularx}
\usepackage{wrapfig}
\usepackage{booktabs}
\usepackage{lettrine}
\usepackage[a4paper,bottom=20mm,right=20mm,left=20mm,top=20mm]{geometry}
\usepackage[super,comma, sort&compress]{natbib}
\usepackage{float}
\usepackage{indentfirst}
\usepackage[margin=0mm]{caption}
\usepackage{subcaption}
\usepackage[separate-uncertainty = true,multi-part-units=single]{siunitx}
\DeclareSIUnit{\sample}{S}
\allsectionsfont{\normalfont\sffamily\bfseries}

\definecolor{darkblue}{rgb}{0,0,0.5}
\definecolor{darkgreen}{rgb}{0,0.5,0}
\definecolor{darkred}{rgb}{.7,0,0}
\definecolor{purple}{rgb}{0.5,0,0.6}
\definecolor{orange}{rgb}{1,0.5,0}
\definecolor{grey}{rgb}{.6,.6,.6}
\definecolor{lightpink}{rgb}{1,0.7,0.75}
\definecolor{pink}{rgb}{1,0.4,0.58}
\definecolor{deeppink}{rgb}{1,0.08,0.58}

\newcommand{\beginsupplement}{%
        \setcounter{page}{0}
        
        \setcounter{table}{0}
        \renewcommand{\tablename}{SUPPL. TAB.}%
        \setcounter{figure}{0}
        \renewcommand{\figurename}{SUPPL. FIG.}
        \setcounter{equation}{0}
        \renewcommand{\theequation}{S.\arabic{equation}}
        \sectionfont{\normalfont\sffamily\bfseries\fontsize{12}{15}\selectfont}
     }

\begin{document}
\setlength{\abovedisplayskip}{8pt} % space above equation
\setlength{\belowdisplayskip}{8pt} % space below equations
% \begin{multicols}{2}
% [ % multicol
\begin{center}
    %%% T I T L E %%%
    %
		\textsf{\textbf{\LARGE In-flight distribution of an electron within\\ a surface acoustic wave}}\\\vspace{2mm}
		
    {\small
		Hermann Edlbauer$^{1,\dagger}$,
        Junliang Wang$^{1,\dagger}$, % GOT FEEDBACK
        Shunsuke Ota$^{2,3,\dagger}$,
        Aymeric Richard$^{1}$,
        Baptiste Jadot$^{1,4}$,\\
        Pierre-Andr\'{e} Mortemousque$^{1,4}$, % GOT FEEDBACK
        Yuma Okazaki$^{3}$, % GOT FEEDBACK
        Shuji Nakamura$^{3}$,
        Tetsuo Kodera$^{2}$,
        Nobu-Hisa Kaneko$^{3}$,\\ % GOT FEEDBACK
		Arne Ludwig$^{5}$,
		Andreas D. Wieck$^{5}$,
        Matias Urdampilleta$^{1}$,
		Tristan Meunier$^{1}$,
		Christopher B\"auerle$^{1}$ \&\\
		Shintaro Takada$^{3,\star}$
	}
\end{center}
\vspace{-2mm}
\def\einr{2mm}
\def\spazi{-1.7mm}
{\small
\begin{nolinenumbers}
	\-\hspace{\einr}$^1$ Univ. Grenoble Alpes, CNRS, Grenoble INP, Institut N\'eel, 38000 Grenoble, France\\
	\-\hspace{\einr}$^2$ Department of Electrical and
Electronic Engineering, Tokyo Institute of Technology, Tokyo 152-8550, Japan \\
	\-\hspace{\einr}$^3$ National Institute of Advanced Industrial Science and Technology (AIST), National Metrology Institute of Japan\vspace{\spazi}\\
	\-\hspace{\einr}\-\hspace{3mm} (NMIJ), 1-1-1 Umezono, Tsukuba, Ibaraki 305-8563, Japan\\
	\-\hspace{\einr}$^4$ Univ. Grenoble Alpes, CEA, Leti, F-38000 Grenoble, France\\
	\-\hspace{\einr}$^5$ Lehrstuhl f\"{u}r Angewandte Festk\"{o}rperphysik, Ruhr-Universit\"{a}t Bochum\vspace{\spazi}\\
	\-\hspace{\einr}\-\hspace{3mm}Universit\"{a}tsstra\ss e 150, 44780 Bochum, Germany\\
	\-\hspace{\einr}$^\dagger$ These authors contributed equally to this work\\
	\-\hspace{\einr}$^\star$ corresponding author:
	\href{mailto:shintaro.takada@aist.go.jp}{shintaro.takada@aist.go.jp}\\%\vspace{2mm}\\
\end{nolinenumbers}
}
\vspace{10mm}\vspace{-5mm}\rule{\textwidth}{0.4pt}\vspace{-6mm}
\subsection*{Abstract} %  ABSTRACT-- 250 words for APL 
{\bf\vspace{-3mm}
Surface acoustic waves (SAW) have large potential to realize quantum-optics-like experiments with single flying electrons employing their spin or charge degree of freedom. 
For such quantum applications, highly efficient trapping of the electron in a specific moving quantum dot (QD) of a SAW train plays a key role.
Probabilistic transport over multiple moving minima would cause uncertainty in synchronisation that is detrimental for coherence of entangled flying electrons and in-flight quantum operations.
It is thus of central importance to identify the device parameters enabling electron transport within a single SAW minimum.
A detailed experimental investigation of this aspect is so far missing.
Here we fill this gap by demonstrating time-of-flight measurements for a single electron that is transported via a SAW train between distant stationary QDs.
Our measurements reveal the in-flight distribution of the electron within the moving acousto-electric quantum dots of the SAW train.
Increasing the acousto-electric amplitude, we observe the threshold necessary to confine the flying electron at a specific, deliberately chosen SAW minimum.
Investigating the effect of a barrier along the transport channel, we also benchmark the robustness of SAW-driven electron transport against stationary potential variations.
Our results pave the way for highly controlled transport of electron qubits in a SAW-driven platform for quantum experiments. 

\vspace{-3mm}\rule{\textwidth}{0.4pt}\vspace{-8mm}
}
\def\baselinestretch{1.5}\selectfont
% ] % multicol
% \multicollinenumbers

\vspace{10mm}

% paragraph MOTIVATING this work:
The use of sound enables nanoelectronic implementations that often resemble quantum-optics experiments within an original acousto-electric solid-state framework.
A prominent example of this development is surface-acoustic-wave (SAW) technology, which is well-established in consumer-electronics industry and currently celebrates its revival in quantum science\cite{Ford2017, Bauerle2018, Delsing2019}.
The acousto-electric medium can serve for instance as a resonator \cite{Gustafsson2014,Manenti2017,Andersson2019} or as a messenger\cite{Satzinger2018,Bienfait2019} of quantum information from superconducting qubits.
However, a SAW can also be employed to transport a charged quantum system as a whole what is particularly interesting for semiconductor-qubit architectures\cite{Vandersypen2017} and experimental investigations of quantum nonlocality with Fermions\cite{Bertoni2000}.
The acousto-electric potential modulation of the SAW drags a single electron with its quantum properties from one surface-gate defined quantum dot (QD) through a transport channel to another QD \cite{Hermelin2011,McNeil2011,Bertrand2016}.
The SAW-driven transport technique allows high transfer efficiency $P>99\% $ even in single-electron circuits of coupled transport channels\cite{Takada2019}.
Owing to this robustness, the acousto-electric method is capable to shuttle spin-entangled electron pairs between distant QDs without significant additional decoherence\cite{Jadot2021}.
The exact transport process remains however an aspect that is yet not fully understood:
Is the electron loosely surfing on a shallow potential wave or is it well-confined within a specific location of the wave train?

% REVIEW literature on flight-time measurements and guide the reader to our experiment:
To study the physics of electron-transfer techniques, time-resolved measurements have emerged as a useful and reliable tool\cite{Sukhodub2004}.
In the quantum-Hall regime, fast voltage-pulse probing  has been employed to study the interaction of edge states\cite{Kamata2014} and it was applied to demonstrate the dynamics of single electrons emitted from a QD pump\cite{Kataoka2016}.
Recently, time-of-flight measurements have been performed in a similar manner in non-chiral mesoscopic conductors\cite{Arrighi2018}.
For SAW-driven single-electron transport, pulsing techniques have been proposed\cite{Hermelin2011} and were applied to trigger single-electron transport at a specific location of the acousto-electric wave train \cite{Takada2019,Jadot2021}.
However, detailed time-of-flight measurements have not yet been performed.

% Short paragraph providing OUTLOOK AND CONTENT:
Here we present a pulse-probe technique that allows the measurement of the electrons in-flight distribution as it passes with the SAW at a specific point of a transport channel.
To perform the time-of-flight measurement, we first apply a ps-voltage pulse to the reservoir gate (R) of the source QD, which allows us to load the electron into a specific SAW minimum.
Subsequently, we apply another ps-pulse on a barrier gate (\#1 or \#2) that is placed along the transport channel as the SAW train passes with the flying electron. 
Sweeping the delay of this probe event, we scan the presence of the electron and map its in-flight distribution for each SAW minimum.
By changing the acousto-electric amplitude we estimate the threshold necessary to confine the electron in the initially loaded SAW minimum.
Introducing a local surface-gate controlled barrier along the transport channel, we investigate further the effect of a stationary potential variation on the in-flight distribution and compare the experimental results to potential simulations.
The present time-resolved investigation provides an important benchmark for the acousto-electric amplitude and the potential homogeneity that is required to perform robust SAW-driven qubit transport and in-flight quantum operations.

\begin{figure}[!h]
    \centering
    \includegraphics[width=6.69in]{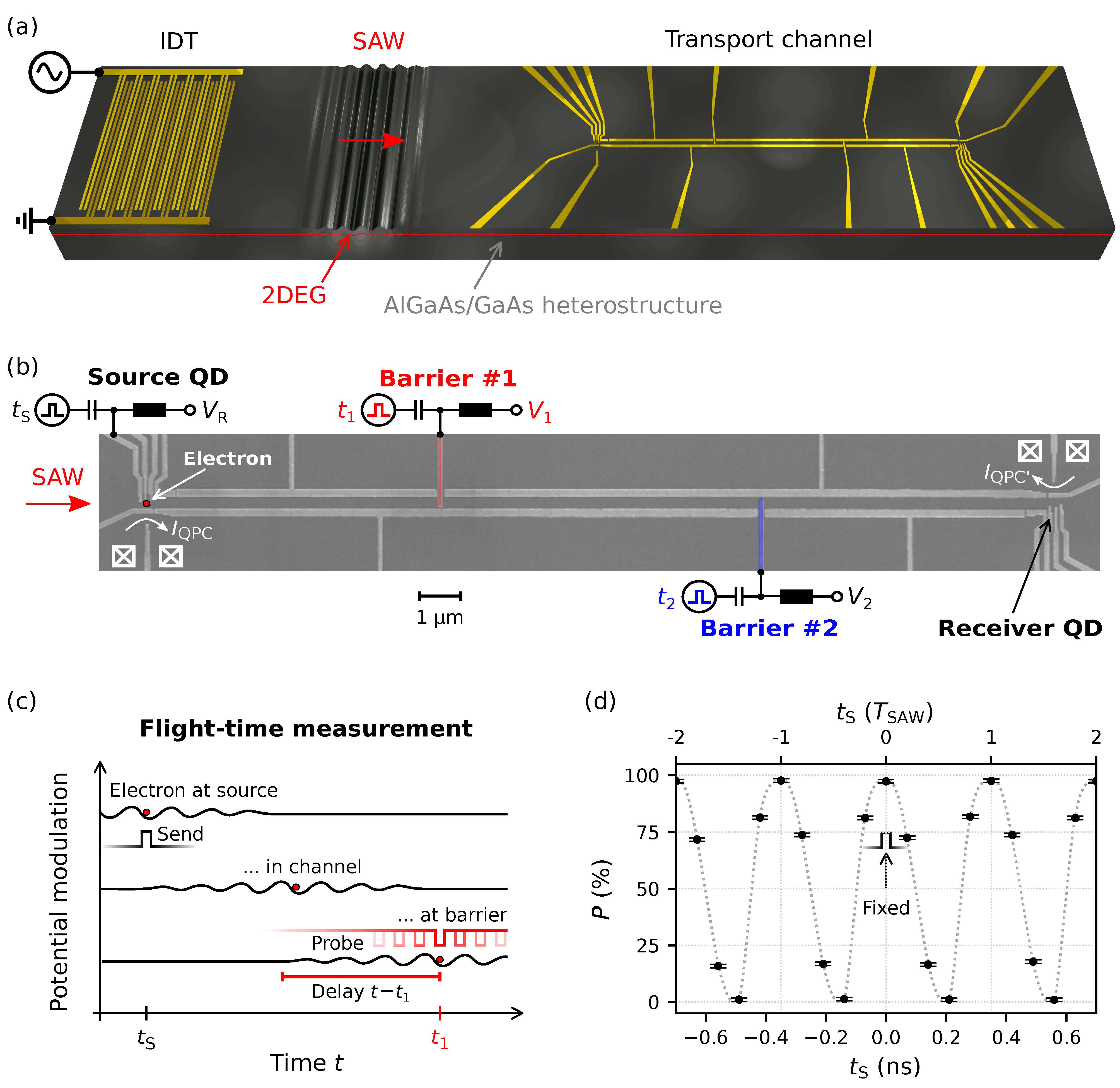}
	\caption{
		\protect
		\textsf{\textbf{Experimental setup.}}
		\textsf{\textbf{(a)}}
        Schematic the AlGaAs/GaAs sample showing an interdigital transducer (IDT) that launches a SAW train towards the surface-gate defined transport channel.
		\textsf{\textbf{(b)}}
		SEM image of transport channel showing source and receiver quantum dots (QD) with schematic indication of quantum-point-contact (QPC) charge detectors and the voltage-pulse to perform the time-of-flight measurement.
		\linebreak\textsf{\textbf{(c)}}
		Schematics showing principles of the controlled sending process at the source and principle of the time-of-flight measurement.
		\textsf{\textbf{(d)}}
		Transfer probability, $P$, as function of the sending pulse $t_\textrm{S}$.
		The probability values $P$ are obtained from $N=13\;000$ single-shot transmissions with error $\Delta P = 1/\sqrt{N}$.
		The dotted line serves as guide to the eye.
		The arrow indicates the delay of the sending pulse that is kept fixed for the time-of-flight measurements.
		\label{fig:setup}
	}
\end{figure}

% Describe the general setup:
The experiment is performed at a temperature of about \SI{20}{mK} within a $^3$He/$^4$He dilution refrigerator.\linebreak
A Si-modulation-doped GaAs/AlGaAs heterostructure serves as basis of the investigated sample that is schematically shown in Fig. \ref{fig:setup}a.
The two-dimensional electron gas (2DEG) that is located \SI{110}{nm} below the surface has an electron density of $n\approx$ \SI{2.8e-11}{cm^{-2}} and a mobility of $\mu\approx$ \SI{9e5}{cm^{2}V^{-1}s^{-1}}.
The main component of the sample is a surface-gate-defined transport channel whose ends are equipped with a quantum dot (QD) serving respectively as single-electron source and receiver.
A scanning-electron-microscopy (SEM) image of the transport channel is shown in Figure \ref{fig:setup}b with schematics indicating the experimental setup.
The Schottky gates of the nanostructure are made out of a metal stack of \SI{3}{nm} titanium and \SI{14}{nm} gold.
During the cool down a voltage of $\SI{0.3}{V}$ is applied to all electrodes.
At low temperature, we completely deplete the 2DEG lying below the \SI{20}{\micro m}-long transport channel and control the source and receiver QD via a set of negative voltages.
The occupation of each QD is sensed via the current flowing through an adjacent quantum point contact (QPC).

% Explain role of IDT and the pulsing setup:
An interdigital transducer (IDT) is placed far to the left of the nanostructure which allows single-shot emission of a SAW train -- with \SI{1}{\micro m} wavelength -- that propagates with a speed of $v_\textrm{SAW}\approx$ \SI{2.86}{\micro m\per ns} towards the transport channel\cite{idt}.
The input signal for the IDT stems from an output channel of an arbitrary waveform generator (AWG).
In order to shuttle a single electron from one QD to the other, we employ the potential modulation that accompanies the SAW in the piezoelectric substrate.
At the depleted transport channel, this acousto-electric modulation forms a train of potential minima that is capable to move the electron.
% Talk about pulsing setup:
We trigger the SAW-driven sending event via the reservoir gate (R) of the source QD that is connected to an AWG output.
To carry out the time-of-flight measurement, additionally, two barriers (\#1 and \#2) are placed along the channel that are also linked to AWG outputs.
Details to the employed IDT and the microwave setup are provided in supplementary sections 1 and 2.

% Explain the transport procudure:
Before the emission of the SAW train, an electron is loaded at the source QD and prepared in a protected configuration where the acousto-electric wave cannot pick up the electron.
In this sending configuration a slight potential variation is however sufficient to move the electron from the stationary source QD into one of the moving QDs that accompanies the SAW train along the transport channel -- see red schematic in Figure \ref{fig:setup}c.
At the same time, we prepare the receiver quantum dot in a configuration where we can catch the moving electron.
Details to the calibration of the SAW-driven transport procedure are provided in supplementary section 3. 
Figure \ref{fig:setup}d shows the transfer probability as function of the source-pulse delay, $t_\textrm{S}$.
As in previous investigations\cite{Takada2019}, we find that $t_\textrm{S}$ must coincide with a specific acousto-electric pressure phase of the SAW in order to enable the electron's transfer from the stationary source QD into a specific location within the SAW.

% Describe how the flight-time measurement works:
To perform the time-of-flight measurement, we fix the timing of the sending trigger ($t_\textrm{S}=$ \SI{0}{ns}) and sweep the delay of a probe pulse $t$ at the according barrier over the arrival window of the SAW.
We have chosen the probe-pulse width as two fifths of the SAW period $T_{\rm SAW}$ in order to optimize pulsing efficiency and time resolution.
The delay of the probe pulse is stepped in multiples of the period $T_\textrm{SAW}$ with a time offset enabling coincidence with the pressure phase of the SAW.
If the probe pulse overlaps with a certain SAW minimum, it enables thus a potentially present electron to pass.
Otherwise, the passage of the electron is blocked leaving the electron behind the barrier in the subsequent SAW period.
Let us first consider the hypothetical case where the electron is well-confined in a specific SAW minimum.
In this case, the transfer probability is zero for a barrier-pulse delay before the electron arrival.
For any delay in the pressure phase after the arrival, the electron is on the other hand certainly transmitted, since it is blocked before the barrier until the probe pulse is present.
Following this measurement approach, the instantaneous distribution of the electron within the passing SAW train, $D(t)$, is directly reflected via the derivative of the transfer probability, $P$, with respect to the time-delay, $t$, of the probe pulse if the effect of the barrier pulse is fully deterministic (block or transmit).
We estimate the uncertainty in $D(t)$ via error propagation as approximately three times the error of $P=1/\sqrt{N}$ where $N$ indicates the number of single-shot transfers.
For a deterministic probe pulse and single-minimum transport, $D(t)$ should follow a delta function at the SAW location expected from the triggered sending event ($t_{\rm S}$).

\begin{figure}[!h]
	\begin{minipage}[c]{.48\textwidth}
    	\caption{
    		\protect
    		\textsf{\textbf{Time-of-flight measurement at barrier \#1.}}
    		\textsf{\textbf{(a)}}
    		Normalized transfer efficiency $\hat P$ as function of delay on the probe pulse $t$.
    	    Each data point is derived from $N=25\;000$ single-shot transmissions with error $\Delta P = 1/\sqrt{N}$.
    	    The dotted line shows a step function centered at the expected arrival time $t_1$.
            \textsf{\textbf{(b)}}
            In-flight distribution $D(t)$ of the electron within the SAW train for different values of the peak-to-peak SAW amplitude $A_\textrm{SAW}$.
            The data is obtained via the normalized derivative of $P(t)$ with uncertainty $\Delta D \approx 3\cdot\Delta P$.
            The arrows indicate the confinement fidelity $D_0 = D(t_1)$.
    		\textsf{\textbf{(c)}}
    		$D_0$ as function of SAW amplitude $A_\textrm{SAW}$.
    		The correspondence between SAW amplitude and input power is provided in supplementary section 4.
    		The dotted curves show a guide to the eye (sigmoid function) and the 95\% threshold.
    		The two vertical lines indicate the SAW amplitude employed in a previous work\cite{Takada2019} and the amplitude for significant confinement.
    % 		The two vertical dotted lines indicate the SAW amplitude employed in a previous work\cite{Takada2019} and the 95\% localization threshold.
    % 		In addition, an exponential fit to the data is shown serving as guide to the eye.
    % 		\comment{CB: what is the meaning of the red arrow in figure d) ,.}
    		\label{fig:tofsaw}
    	}
    \end{minipage}
    \hfill
	\begin{minipage}[c]{.5\textwidth}
    	\centering
        \includegraphics[width=\textwidth]{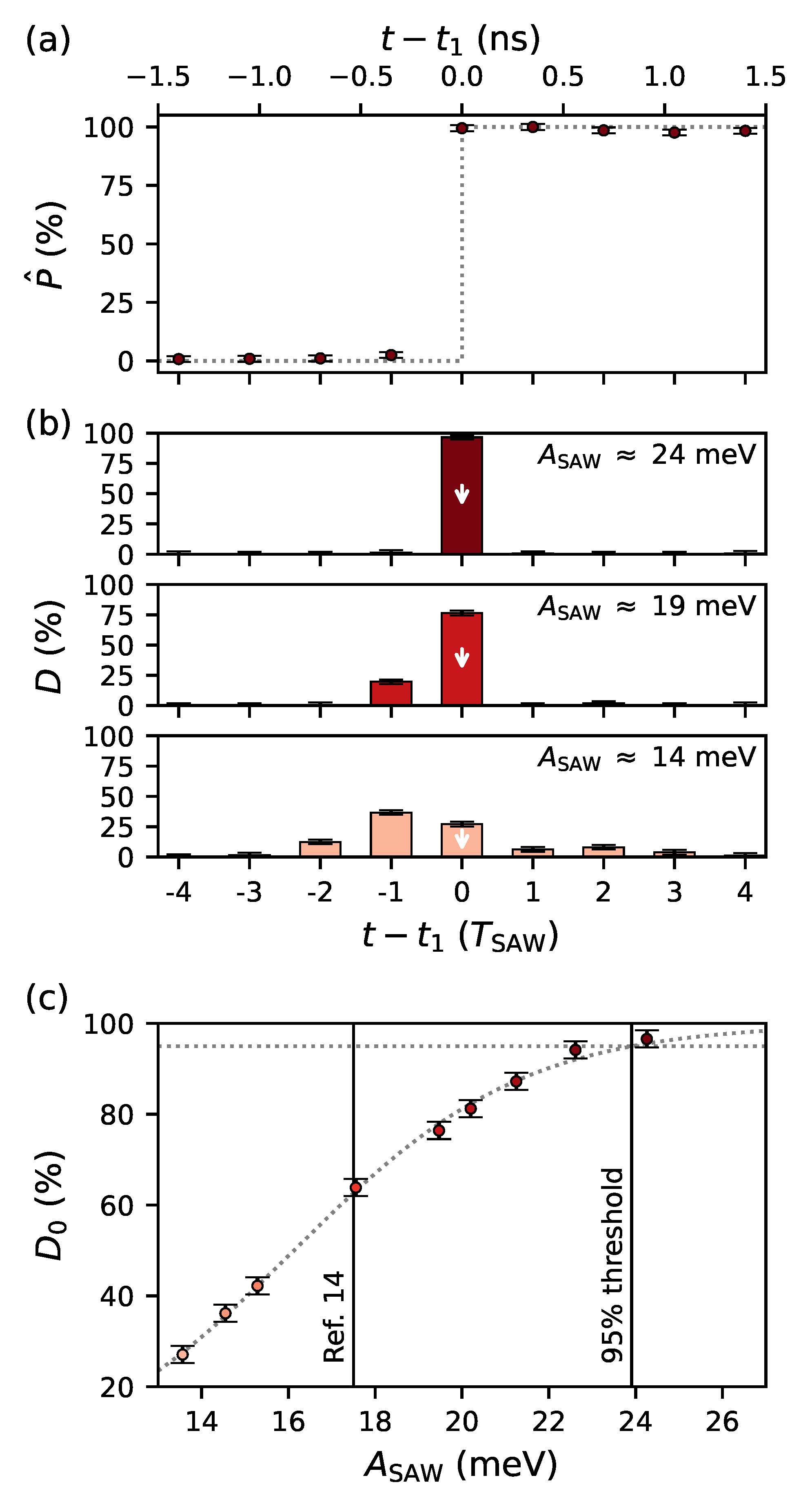}
    \end{minipage}
\end{figure}

% Describe flight-time data for maximum SAW power and draw conclusions:
Let us first perform the time-of-flight measurement at the maximum achievable IDT input power of about \SI{288}{mW} (24.6 dBm).
Based on SAW-modulated Coulomb-peak data\cite{Takada2019,Schneble2006} -- see supplementary section 4 --, we estimate that this input power introduces a SAW with amplitude $A_{\rm SAW}\approx (24\pm3)\;\textrm{meV}$.
We estimate the expected arrival time ($t_1$) at barrier \#1 from its distance to the source QD ($x_1$) as
$
    t_1 = 
    x_1/v_\textrm{SAW} \approx 
    (2.23\pm0.05)\;\textrm{ns}
$.
Sweeping the delay of the probe pulse on barrier \#1 over the time window around $t_1$, we measure the transfer probability $P(t)$ as shown in Fig. \ref{fig:tofsaw}a.
The data follows a step function that is centered at $t_1$ what verifies the deterministic effect of the barrier pulse and transport in a specific SAW position.
It is important to note, that the binary probing process is achieved via the sample design allowing a spatial separation of the probing barrier to the catching event at the receiver QD.
Evaluating the electron distribution $D(t)$, we find that the electron is confined during transport with a certainty of $(97\pm2)\%$ at the position of the SAW train ($t-t_1=0$), where the electron was initially loaded with the ps-voltage pulse.
We suspect that even more confinement is achievable if the SAW amplitude would be further increased.

% Describe how reduced SAW amplitude affects the transfer distribution:
In order to investigate this aspect in more detail, let us focus on the threshold of SAW amplitude below which confinement breaks down.
For this purpose, we repeat the time-of-flight measurement at the first barrier with gradually decreasing SAW amplitude as shown in Fig. \ref{fig:tofsaw}b.
At a SAW amplitude of about 19 meV, we observe partial electron arrival one SAW minimum earlier than expected ($t-t_1=-T_\textrm{SAW}$).
For a SAW amplitude below 15 meV, we find the arrival distribution also spreading over SAW minima subsequent to the expected location.
We quantify the magnitude of electron displacement by plotting $D_0 = D(t_1)$ as function of the SAW amplitude as shown in Fig. \ref{fig:tofsaw}c.
The data shows significant ($D_0 > 95\%$) in-flight confinement of the transported electron for a SAW amplitude exceeding a threshold of $(24\pm3)\;\textrm{meV}$.
The course of the data indicates that increased SAW amplitude indeed allows maximization of in-flight confinement.

\begin{figure}[!h]
	\begin{minipage}[c]{.48\textwidth}
    	\caption{
    		\protect
    		\textsf{\textbf{Time-of-flight measurement at barrier \#2.}}
            \textsf{\textbf{(a)}}
            In-flight distribution $D(t)$ of the electron within the SAW-train for selected voltages $V_1$ applied on the first barrier gate.
            The measurement is performed with maximum SAW amplitude of about \SI{24}{meV}.
    	    Each data point is derived from $N=3\;000$ single-shot transmissions with undertainty $\Delta D \approx 3/\sqrt{N}$.
    	    \textsf{\textbf{(b)}}
    		Confinement fidelity $D_0 = D(t_1)$ for different values of $V_1$.
            The dotted curves show a guide to the eye (polynomial fit) and the 80\% threshold. 
             \textsf{\textbf{(c)}}
             Minimum $U_\textrm{min}$ of the electron potential across the first barrier three selected voltages: \SI{-217}{mV}, \SI{-121}{mV} and \SI{-25}{mV}.
             The dotted curve shows the potential modulation of the applied SAW for comparison with indication of its peak-to-peak amplitude $A_\textrm{SAW}$.
    		 \label{fig:tofpot}
    	}
    \end{minipage}
    \hfill
	\begin{minipage}[c]{.5\textwidth}
    	\centering
        % EFFECT OF POTENTIAL LANDSCAPE
        \includegraphics[width=\textwidth]{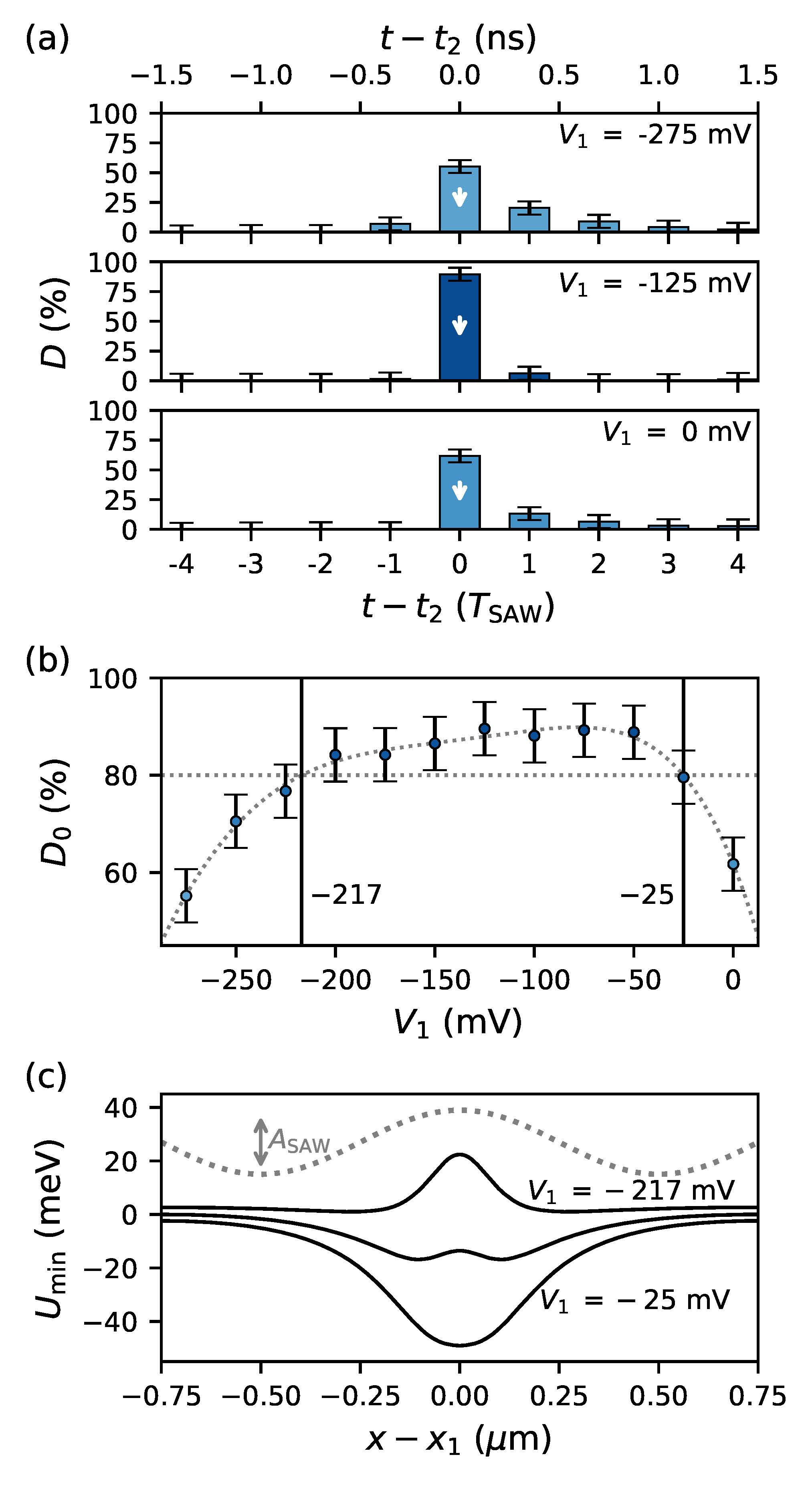}
    \end{minipage}
\end{figure}

% Discuss the effect of a potential barrier:
The presence of two controllable barriers along the transport channel allows to study the effect of an intermediate potential variation on the time-of-flight distribution.
For this purpose, we probe the electron's arrival at the second barrier.
There, we expect the electrons passage at a  time of
$
    t_2 = 
    x_2/v_\textrm{SAW} \approx 
    (4.86\pm0.05)\;\textrm{ns}
$
where $x_2$ indicates the distance of barrier \#2 to the source QD.
On the first barrier we do not apply a voltage pulse, but only change the static voltage $V_1$.
Figure \ref{fig:tofpot}a shows data of such a time-of-flight measurement with intermediate barrier.
Here, we keep the SAW amplitude at the maximum achievable value.
Despite the presence of the intermediate barrier, the data shows a confinement fidelity $D_0>80\%$ in the voltage range between $\SI{-217}{mV}$ to $\SI{-25}{mV}$.
Beyond this range, we observe a rapid drop of $D_0$ due to in-flight transitions of the electron in neighboring SAW minima.
The presence of the intermediate barrier causes a reduction of the confinement fidelity compared to a smooth potential along the transport channel.

% Discuss the data via potential simulations: 
To understand the experimental result better, let us compare the time-of-flight data to potential simulations along the transport channel.
The potential calculation is performed via the commercial Poisson solver nextnano\cite{Birner2007} assuming a frozen charge layer and deep boundary conditions\cite{Hou2018}.
As input parameters, we consider the gate geometry and the voltages applied as well as the intrinsic properties of the heterostructure.
Figure \ref{fig:tofpot}c shows the course of the resulting potential minimum $U_{\rm min}$ across the first barrier for three values of the voltage $V_1$ beyond the 80\%-confinement threshold.
The blockade of the electron for a very high ($V_1<-217\;\textrm{mV}$) and very low ($V_1>-25\;\textrm{mV}$) potential barrier is caused by different effects.
For increasingly negative barrier voltage, a potential barrier is formed.
Interestingly, the formation of the barrier comes along with the event, where the electron arrives one SAW minimum earlier as apparent in Fig. \ref{fig:tofpot}a for $V_1=-275\;\textrm{mV}$ at $t-t_2=-1\cdot T_{\rm SAW}$.
We speculate that the advancement is caused via non-adiabatic transitions of the electron in excited states when ramping against the barrier.
For increasingly positive voltage such as $V_1=0\;\textrm{mV}$, we observe on the other hand mainly transitions in subsequent SAW minima ($t-t_2>0$) due to trapping in a quantum-dot like potential structure.
The time-of-flight measurement at the second barrier shows that single-minimum transport is feasible even if significant potential gaps are present along the transport channel.
We suspect that maximized acousto-electric confinement (via increased SAW-amplitude and -frequency) enables single-electron transfer that is fully protected against local potential variations \cite{Nixon1990}.

The ability to detect the exact location of a SAW-transported electron is essential to identify the critical device parameters enabling single-electron transport within a specific, deliberately chosen SAW minimum and thus flying-qubit implementations.
Employing a pulse-probe technique, we have demonstrated time-of-flight measurements revealing the arrival distribution of such a flying electron for each moving potential minima that accompanies the SAW along a transport channel.
As we send a SAW train with sufficiently large peak-to-peak amplitude -- that is $(24\pm3)\;\textrm{meV}$ for the presently investigated device --, the time-of-flight data indicates an electron-confinement fidelity exceeding 95\% within a specific moving QD.
Investigating the effect of an intermediate surface-gate-defined barrier within the transport channel, we demonstrate a confinement fidelity larger than 80\% over a voltage range of about \SI{200}{mV} despite the presence of the local potential variation.
Our measurements show that acousto-electric in-flight confinement plays the key role to make a SAW-driven single-electron circuit robust.
We anticipate that our experimental findings foster the development and application of SAW-generation approaches \cite{Buyukkose2012,Schulein2015,Ekstrom2017,Dumur2019,Nysten2020,Weiss2018} paving the way for SAW-driven electron-quantum-optics implementations \cite{Barnes2000,Shukur2017,Schuetz2017,Lepage2020} and quantum-metrology applications\cite{Cunningham2000}.

\subsection*{Supplementary material}
See supplementary material for detailed description of 
(1) the IDT and the employed SAW signal,
(2) the setup for the voltage pulses,
(3) single-electron-transport procedure and
(4) the estimation of the acousto-electric potential modulation.

 \subsection*{Data availability}
\vspace{12pt}
The datasets used in this work are available from the corresponding author on reasonable request.
\vspace{12pt}

\subsection*{Acknowledgements}
J.W. acknowledges the European Union's Horizon 2020 research and innovation program under the Marie Skłodowska-Curie grant agreement No 754303.
T.K. and S.T. acknowledge financial support from JSPS KAKENHI Grant Number 20H02559.
N.-H.K. acknowledges financial support from JSPS KAKENHI Grant Number JP18H05258.
C.B. acknowledges funding from the European Union’s H2020 research and innovation programme under grant agreement No 862683. 
% and from ANR-project "QUABS". 

\pagebreak

\subsection*{References}
\def\bibsection{}

\pagebreak

\pagebreak

\pagebreak

\pagebreak
\beginsupplement

\;\vfill
\begin{center}
	\textsf{\textbf{\Huge SUPPLEMENTARY MATERIAL\\\vspace{2mm}
	\Large In-flight distribution of an electron within\\ a surface acoustic wave}}\\\vspace{2mm}
		
    {\small
		Hermann Edlbauer$^{1,\dagger}$,
        Junliang Wang$^{1,\dagger}$,
        Shunsuke Ota$^{2,3,\dagger}$,
        Aymeric Richard$^{1}$,
        Baptiste Jadot$^{1,4}$,\\
        Pierre-Andr\'{e} Mortemousque$^{1,4}$,
        Yuma Okazaki$^{3}$,
        Shuji Nakamura$^{3}$,
        Tetsuo Kodera$^{2}$,
        Nobu-Hisa Kaneko$^{3}$,\\
		Arne Ludwig$^{5}$,
		Andreas D. Wieck$^{5}$,
        Matias Urdampilleta$^{1}$,
		Tristan Meunier$^{1}$,
		Christopher B\"auerle$^{1}$ \&\\
		Shintaro Takada$^{3,\star}$
	}
\end{center}
\vspace{-2mm}
\def\einr{2mm}
\def\spazi{-1.7mm}
{\small
\begin{nolinenumbers}
	\-\hspace{\einr}$^1$ Univ. Grenoble Alpes, CNRS, Grenoble INP, Institut N\'eel, 38000 Grenoble, France\\
	\-\hspace{\einr}$^2$ Department of Electrical and
Electronic Engineering, Tokyo Institute of Technology, Tokyo 152-8550, Japan \\
	\-\hspace{\einr}$^3$ National Institute of Advanced Industrial Science and Technology (AIST), National Metrology Institute of Japan\vspace{\spazi}\\
	\-\hspace{\einr}\-\hspace{3mm} (NMIJ), 1-1-1 Umezono, Tsukuba, Ibaraki 305-8563, Japan\\
	\-\hspace{\einr}$^4$ Univ. Grenoble Alpes, CEA, Leti, F-38000 Grenoble, France\\
	\-\hspace{\einr}$^5$ Lehrstuhl f\"{u}r Angewandte Festk\"{o}rperphysik, Ruhr-Universit\"{a}t Bochum\vspace{\spazi}\\
	\-\hspace{\einr}\-\hspace{3mm}Universit\"{a}tsstra\ss e 150, 44780 Bochum, Germany\\
	\-\hspace{\einr}$^\dagger$ These authors contributed equally to this work\\
	\-\hspace{\einr}$^\star$ corresponding author:
	\href{mailto:shintaro.takada@aist.go.jp}{shintaro.takada@aist.go.jp}\\%\vspace{2mm}\\
\end{nolinenumbers}
}
\vfill
\tableofcontents
\thispagestyle{empty}
\vfill
\pagebreak

\section{Interdigital transducer}
\label{suppl:idt}

\begin{figure}[!b]
    \centering
    \includegraphics[width=\textwidth]{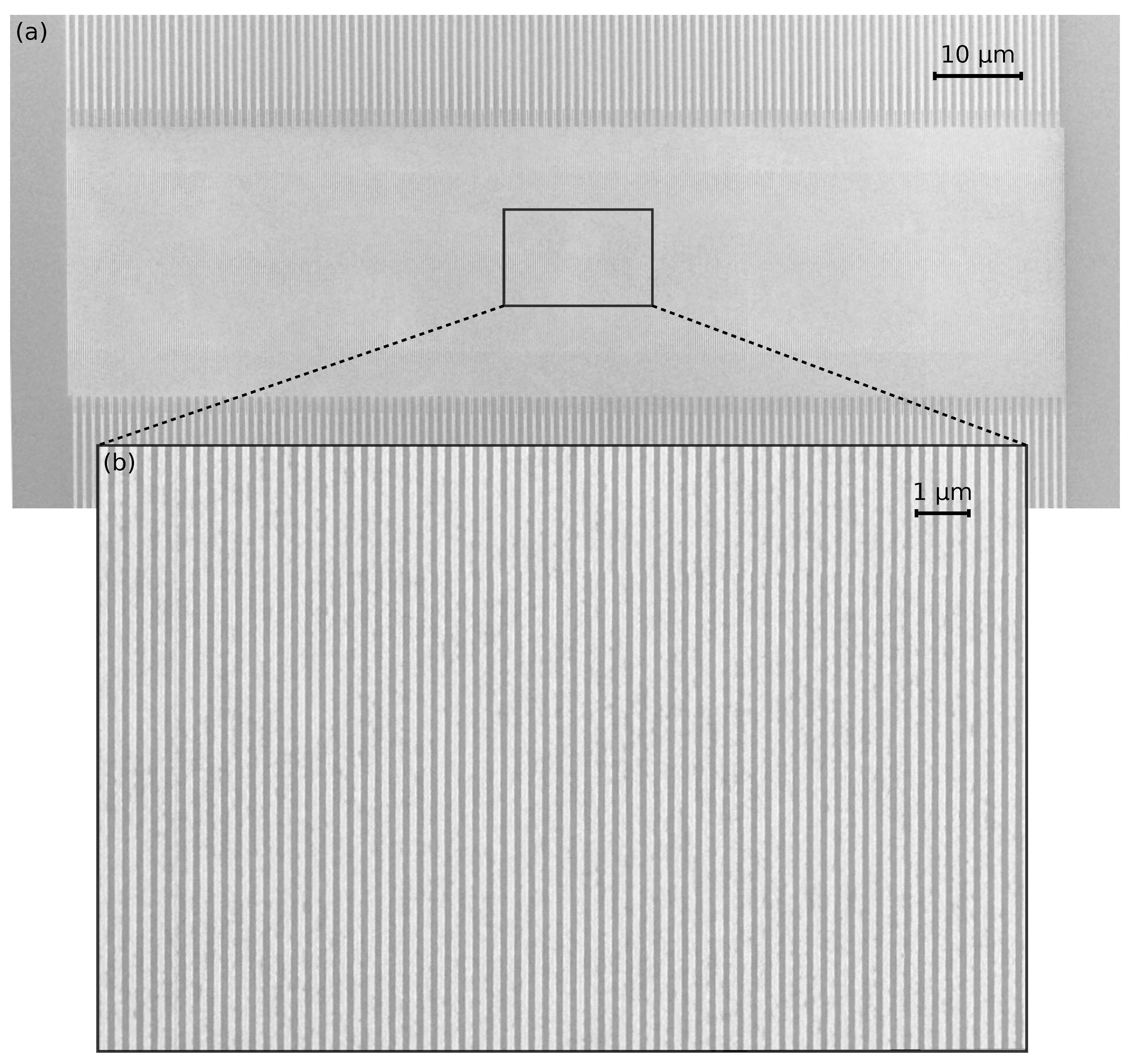}
	\caption{
		\textsf{\textbf{SEM image of the employed interdigital transducer.}}
        \textsf{\textbf{(a)}} Overview showing whole double-finger IDT.
        \textsf{\textbf{(b)}}
        Zoom into IDT center showing the homogeneity of the interlocked double-finger 
        electrodes.
	}\label{sfig:idt}
\end{figure}

The employed double-finger IDT is fabricated using state-of-the-art electron-beam lithography.
A SEM image of the IDT electrodes of the presently investigated device is shown Suppl. Fig. \ref{sfig:idt}.
After development of the electron-beam resist, a stack of \SI{3}{\nano \meter} titanium and \SI{27}{\nano \meter} aluminium is deposited via thin-film evaporation.
The IDT is positioned at a distance of about \SI{1.6}{\milli \meter} left to the nanodevice and consists out of 111 interlocked double-finger electrodes having a periodicity of \SI{1}{\micro m}.
The IDT's length corresponds thus to a SAW flight time of about \SI{39}{ns}.
The aperture of the IDT fingers is \SI{30}{\micro m} and the metallization ratio close to 1/2.
To launch a SAW train, we apply a resonant signal during a period of \SI{50}{ns} on one of the IDT electrodes -- the other is connected to cold ground.
In turn, a 89-ns-long SAW train is emitted in a single shot that propagates with a speed of \SI{2.86}{\micro \meter \per \nano \second} along the depleted channel.
The SAW train shows a trapezoidal envelope with a central amplitude saturation region of about 11 ns what corresponds to 30 SAW periods.
\pagebreak

\section{Voltage-pulse setup}
\label{suppl:rfsetup}

\begin{figure}[!h]
	\begin{minipage}[c]{.4\textwidth}
	    \caption{
		\textsf{\textbf{Voltage-pulse setup.}}
        Schematic showing the cabling of the four transmission lines from the arbitrary waveform generator (AWG) to the interdigital transducer (IDT) and the three pulsing gates of the nanodevice going to the source quantum dot (S) and the two barriers (1 and 2) placed along the transport channel.
        Additional amplification and attenuation is indicated for the different temperature stages of the dilution refrigerator ranging from room temperature (RT) to the base temperature of about \SI{20}{mK}.
        The schematic does not show the electrical connections to the static gates that are not used for pulsing.
		\label{sfig:rfsetup}
	    }
    \end{minipage}
    \hfill
	\begin{minipage}[c]{.55\textwidth}
    	\centering
    	\includegraphics[width=\textwidth]{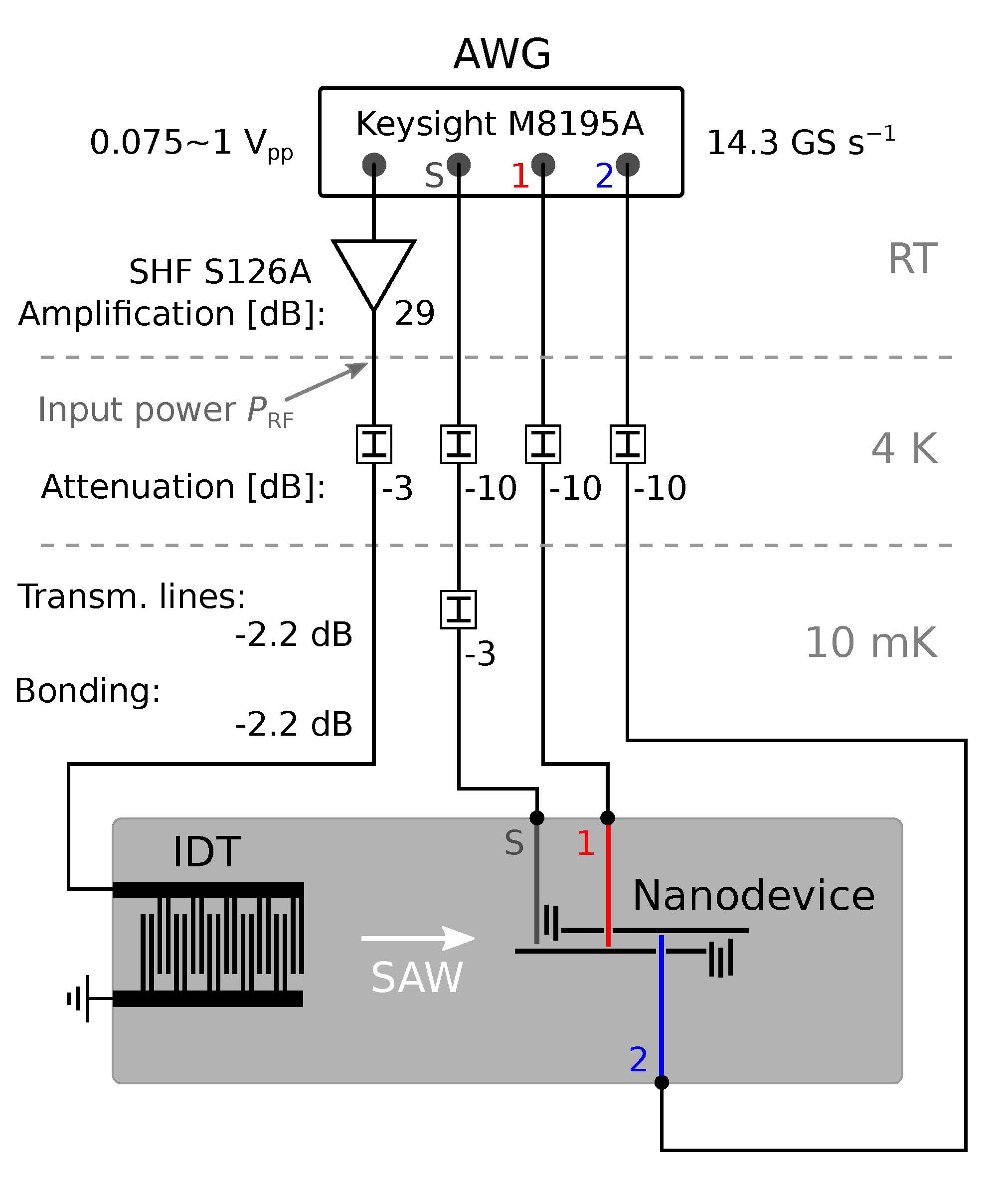}
    \end{minipage}
\end{figure}

The input signal for the IDT and the three pulsing gates are provided via an arbitrary waveform generator (AWG) of the type Keysight M8195A.
A schematic description of the microwave setup is shown in Suppl. Fig. \ref{sfig:rfsetup}.
The AWG is set in sequence mode with extended memory on four output channels.
According to the AWG internal circuit, the sampling rate of \SI{65}{\giga\sample\per\second} is divided by a factor of four.
In order to match a multiple of the SAW resonance frequency, we adjusted this reduced sampling rate to \SI{14.3}{\giga\sample\per\second} that is five times the resonance frequency.
Testing the AWG output on a fast sampling oscilloscope, we verify a jitter below 6 ps between the four output channels.
The AWG output towards the IDT is equipped with an amplifier of the type SHF S126A providing a gain of 29 dB.
The amplifier's input saturation of 25 dBm (for 1 dB compression) sets the upper limit of the input power $P_{\rm RF}$ for SAW generation.
The AWG signal for the IDT was thus limited to a signal with a maximum peak-to-peak amplitude of 400 mV what corresponds to the amplifier's saturation power with subtracted gain.
The transmission lines are equipped with attenuators to mitigate the injection of thermal noise.
The attenuation of the four transmission lines was deduced at a temperature of 4 K via a network analyzer measurement as $-2.2$ dB at the IDTs resonance frequency.
The attenuation of the bonding wires from the printed circuit board (PCB) carrying the GaAs-based device is also estimated as $-2.2$ dB.
The transmission lines of pulsing gates 1, 2 and S are equipped with bias tees at the lowest temperature stage.

\vfill

\vfill
\pagebreak

\section{Single-electron transfer}
\label{suppl:transfer}

\begin{figure}[!h]
    \centering
    \includegraphics[width=6.69in]{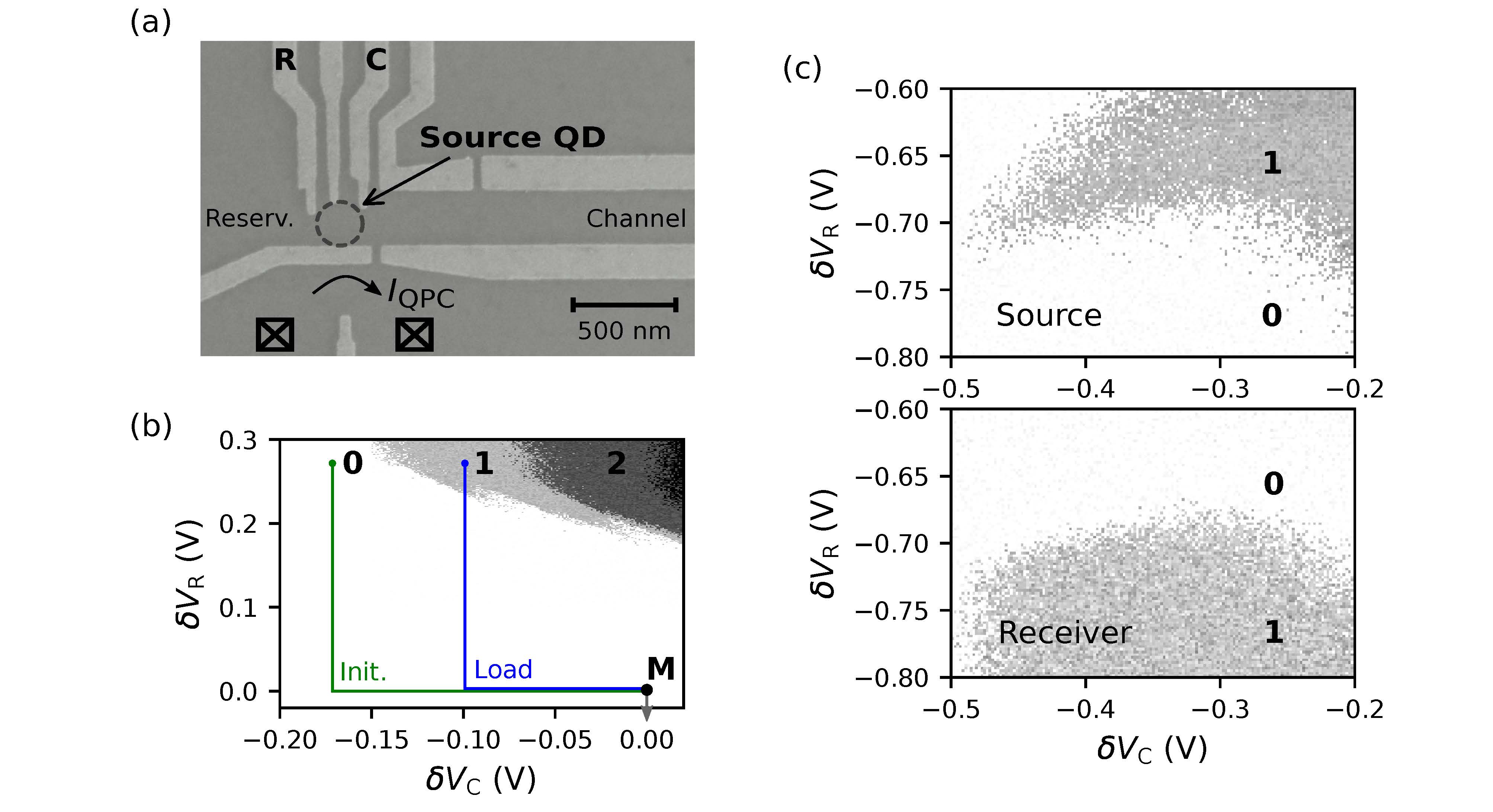}
	\caption{
		\textsf{\textbf{Loading and transfer maps.}}
        \textsf{\textbf{(a)}}
        SEM image of the source QD with schematic indications of the essential controls.
        \textsf{\textbf{(b)}}
        Loading map showing quantum-dot occupation for different configurations of the reservoir gate $V_R$ and the channel gate $V_C$.
        The data shows the difference in QPC current $\Delta I_{\rm QPC}$ after voltage-movements ($\delta V_R$ and $\delta V_R$) from the measurement point (M) to a loading configuration.
        \textsf{\textbf{(c)}}
        Transfer map showing change of the occupation of the source and receiver QD after loading one electron at the source and SAW emission.
        The data shows the change in the respective QPC current $\Delta I_{\rm QPC}$ for different movements ($\delta V_R$ and $\delta V_R$) to a certain sending configuration.
	}\label{sfig:transfer}
\end{figure}

The characterization of the source and receiver quantum dot (QD) is essential to achieve highly efficient SAW-driven single-electron transfer.
The essential controls for the transfer procedure are the QD's surface gates facing the electron reservoir (R) and the depleted channel (C) as indicated in Suppl. Fig. \ref{sfig:transfer}a.
The presence of an electron is traced via the QPC current $I_{\rm QPC}$ that is recorded for both the source and receiver QD during execution of the transfer procedure.
The so-called loading map shown in Suppl. Fig. \ref{sfig:transfer}b shows the steps of the transfer procedure including initialization, loading and sending.
After each step, the QPC current is read out, which enables an accurate measurement of the QD occupation -- see plateaus indicated as 0, 1 and 2.
Having initialized the QD and loaded a single electron, we bring the QD in a sending configuration where the SAW transfers the electron from the source to the receiver.
After each transfer cycle, we send a SAW pulse to clean the transport channel from a potentially remaining electron.
Supplementary Figure \ref{sfig:transfer}c shows a sending map for the exact transfer configuration that was employed for the time-of-flight measurements.
The data shows that the SAW transfers the electron to the receiver for sufficiently negative $V_{\rm R}$.
By comparing the QD occupation before and after transfer, we estimate the errors of initialization, loading and catching as 0.006\%, 0.2\% and 1\%.
We find that the error of transfer is less than 0.01\%.
The high transfer efficiency is reflected in the complementarity of the transfer data at the source and receiver QD.

\vfill

\vfill

\pagebreak

\section{Acousto-electric potential modulation}
\label{suppl:sawampl}

\begin{figure}[!b]
	\begin{minipage}[c]{.47\textwidth}
	    \caption{
		\textsf{\textbf{Measurement of SAW amplitude.}}
        \textsf{\textbf{(a)}}
		Coulomb diamonds formed via one of the employed quantum dots (QD).
		The data shows the absolute derivative of the current, $I$, through the QD with respect to the bias voltage $V_{\rm SD}$ as function of the plunger gate voltage $V_{\rm P}$ and $V_{\rm SD}$.
        \linebreak\textsf{\textbf{(b)}}
		Coulomb-peak broadening introduced via continuous SAW modulation.
        Here we fix $V_{\rm SD}\approx0.2\;\textrm{mV}$.
		The data shows the absolute derivative of the current, $I$, through the QD with respect to the $V_{\rm P}$ as function of $V_{\rm P}$ and the SAW input power $P_{\rm RF}$.
        To avoid unnecessary heating from the continuous input signal, we increase the input power only up to $P\approx0\;\textrm{dBm}$.
        For this measurement we do not employ the AWG but a signal generator of the type Rohde \& Schwarz SMA100B.
        \textsf{\textbf{(c)}}
        SAW amplitude $A_{\rm SAW}$ as function of $P_{\rm RF}$ according to equation \ref{equ:ampl}.
        The grey area indicates the uncertainty region.
		\label{sfig:sawampl}
	    }
    \end{minipage}
    \hfill
	\begin{minipage}[c]{0.5\textwidth}
    	\centering
    	\includegraphics{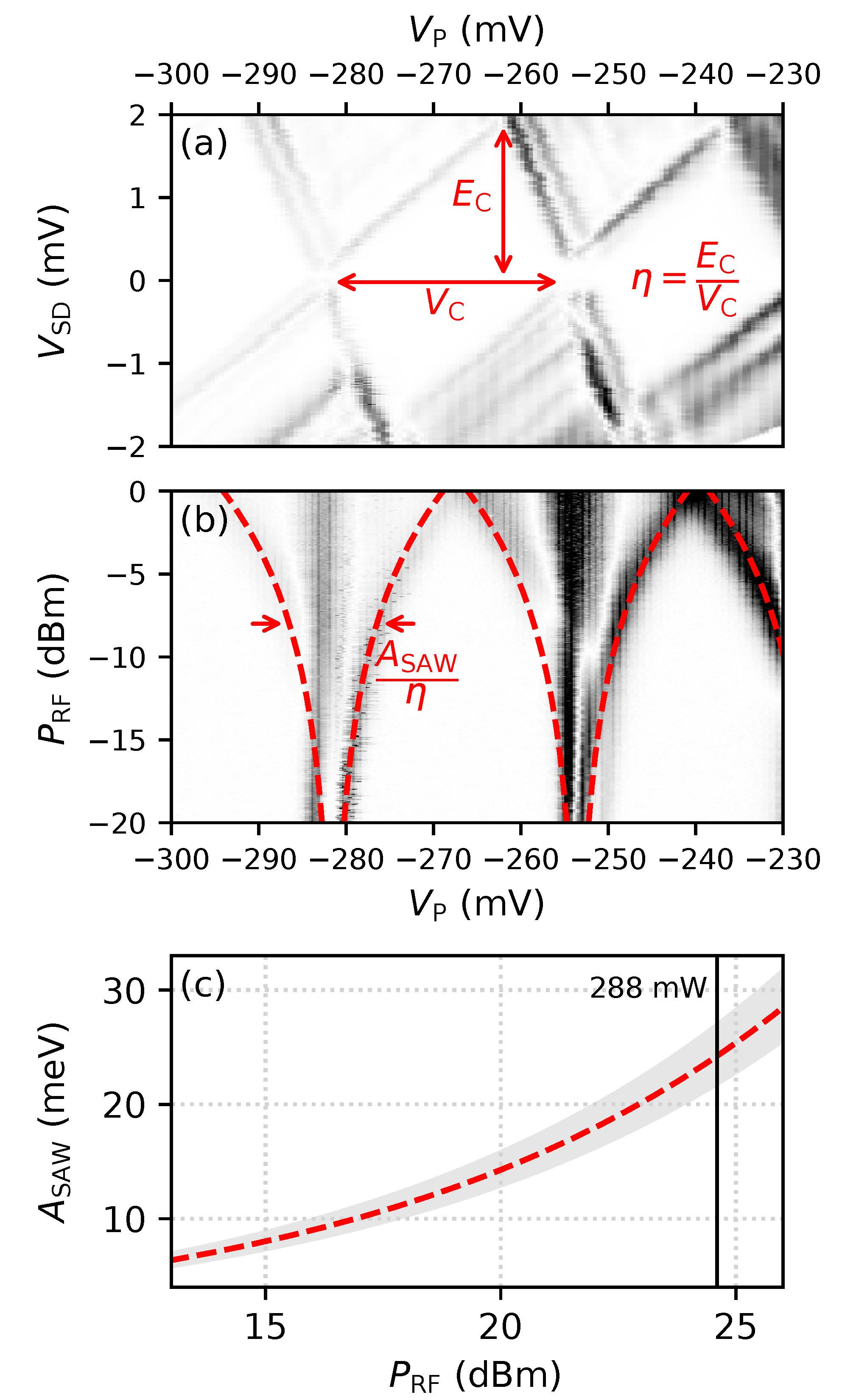}
    \end{minipage}
\end{figure}

In order to deduce the SAW potential modulation from the input power $P_\textrm{RF}$ of a resonant IDT signal, we employ the method of SAW-induced Coulomb-peak broadening.
For this purpose, we couple a QD of our setup to source and drain reservoirs and tune it into the Coulomb-blockade regime.
Varying a small bias $V_{\rm SD}$ across the reservoirs and the plunger-gate voltage $V_{\rm P}$, we measure Coulomb diamonds as shown in Suppl. Fig. \ref{sfig:sawampl}a.
From the ratio of the charging energy $E_{\rm C}$ and the diamond spacing $V_{\rm C}$, we deduce the voltage-to-energy conversion factor as $\eta\approx56.6\;\textrm{meV/V}$ .
Knowing $\eta$, we can now measure the Coulomb peaks while gradually increasing the power $P_{\rm RF}$ as shown in Suppl. Fig. \ref{sfig:sawampl}b.
The Coulomb peaks split according to peak-to-peak amplitude of the SAW following the input-power relation: 
\begin{equation}
A_{\rm SAW} \;\text{[meV]} = 2 \cdot \eta \cdot 10^{(P_{\rm RF}\;\text{[dBm]}-P_0)/20}
\label{equ:ampl}
\end{equation}
Here $P_0$ indicates the total power losses from signal injection into the transmission line to SAW generation.
The red, dashed curves in Suppl. Fig. \ref{sfig:sawampl}b show a fit of equation \ref{equ:ampl} to the data indicating $P_0\approx(38\pm1)\;\textrm{dBm}$.
With this information, we relate the input power $P_{\rm RF}$ to the peak-to-peak potential-modulation amplitude $A_{\rm SAW}$ of the SAW via extrapolation as shown in Suppl. Fig. \ref{sfig:sawampl}c.
For the maximum achievable input power of 24.6\;\textrm{dBm} we deduce a peak-to-peak SAW amplitude of $A_{\rm SAW}\approx(24\pm3)\;\textrm{meV}$.

% \section{Potential simulations}
% \label{suppl:potsim}


\begin{thebibliography}{10}
\bibitem{idt}
	The SAW wavelength $\lambda_{\rm SAW}$ is determined by the period of the metallic fingers of the IDT. For the presently investigated device the SAW wavelength is \SI{1}{\micro m}. From the measured resonance frequency of the IDT, $f_0 =$ \SI{2.86}{GHz}, the SAW velocity is deduced from $v_{\rm SAW} = \lambda_{\rm SAW} \cdot f_0$.
\end{thebibliography}
\end{document}